\documentclass[9pt]{article}
\usepackage{amsfonts}
\usepackage{pstricks}
\usepackage{pst-node}
\usepackage{epsfig}
\usepackage{epsfig}
\usepackage[latin5]{inputenc}
\usepackage[T1]{fontenc}
\usepackage{lipsum}
\usepackage{amsmath}
\usepackage{authblk}

\begin{document}
                \def\ba{\begin{eqnarray}}
                \def\ea{\end{eqnarray}}
                \def\w{\wedge}
                \def\d{\mbox{d}}
                \def\D{\mbox{D}}

\begin{titlepage}
\title{Simple Supergravity pp-Waves }
\author{Tekin Dereli${}^{1,2,}$\footnote{tdereli@ku.edu.tr, tekindereli@maltepe.edu.tr} , Yorgo \c{S}eniko\u{g}lu${}^{1,}$\footnote{yorgosenikoglu@maltepe.edu.tr}}
\date{%
    ${}^{1}$ \small Department of Basic Sciences, Faculty of Engineering and Natural Sciences, \\Maltepe University, 34857 Maltepe,\.{I}stanbul, Turkey\\%
    ${}^{2}$ \small Department of Physics, Ko\c{c} University, 34450 Sar{\i}yer, Istanbul,Turkey\\[2ex]%
    \today
}
\maketitle



\begin{abstract}

\noindent Non-gauge generated exact solutions of simple supergravity field equations that describe pp-waves in Rosen coordinates are presented in the language of complex quaternion valued exterior differential forms.

 \end{abstract}

\vskip 1cm



\end{titlepage}

\newpage

\section{Introduction}
The simple supergravity theory is the locally supersymmetric generalization of Einstein's general theory of relativity\cite{freedman-ferrara-vanN},\cite{deser-zumino}. In a perturbative approach to quantum gravity, the massless spin-2 graviton field that is determined by the metric tensor of space-time is paired with a massless spin-3/2 gravitino field. The classical predictions of ordinary gravity over  long distances
is not expected to get modified by supergravity because of the fermionic nature of the gravitino. Thus the exchange forces due to gravitino can affect mainly the short distance
behavior. Simple supergravity theory is remarkable in several  respects\footnote{See  for instance \cite{salam-sezgin} for an early review.}.  From the point of view of Hamiltonian constraint analysis, it provides the unique "square-root" of the conventional Einstein field equations. Furthermore the causal propagation of higher spin fields in simple supergravity background is secured by local supersymmetry covariance.
Finally the  renormalizability of perturbative supergravity theories improve over that of Einstein's theory of gravitation by several loop orders \cite{deser-kay-stelle}. Another remarkable feature is that the coupled field equations of simple supergravity take on a very simple form in a non-Riemannian space-time where the torsion is fixed algebraically  by a quadratic
expression in the gravitino field\cite{deser-zumino}.Then the motion of spinning test masses along auto-parallels in a supergravity background  needs some further care.

\noindent The $O(N)$ extended supergravity theories had been constructed and various types of  matter couplings had
been achieved very early in the development of supergravity theories \cite{salam-sezgin}.
The requirement of a unique graviton field sets an upper limit $N \leq 8$.
The renormalizability of such generalizations had been investigated over the years to progressively increasing finite loop orders.
This is pushed to considerably high orders in some  recent work \cite{bern et al},
 however, a proof of renormalizability to all orders, if ever going to be achieved, seems to  lie yet far ahead.

\noindent  Various mathematical techniques are used to investigate supergravity theories in general in the literature.
 Supermanifolds with both commuting and anti-commuting coordinates provide one line of approach. On the other hand going to higher dimensional space-times
and then dimensionally reducing down to four dimensions yields unexpected simplifications in the formulation of supergravities. These approaches also provide links between extended supergravity theories and the effective point field theory limits of certain superstring models. The vacua of such theories can be classified by working out exact supergravity solutions. The customary  approach is to consider a family of  exact solutions of the bosonic sector of the coupled field equations with a certain group of isometries. Then given the isometry generators one solves for Killing spinors in the classical background. Thus a sub-family of exact bosonic solutions are made locally supersymmetric and provides an equivalence class  of vacuum solutions. We call such solutions locally super-gauge generated.

\noindent  The exact solutions of  supergravity field equations that are not super-gauge generated in the above sense are sparse.
 A family of exact plane wave solutions of the simple supergravity field equations in 4-dimensions
that include a sub-family of non-gauge generated solutions were first given by Aichelburg and Dereli\cite{dereli-aichelburg}. A gauge related family is also constructed\cite{urrutia}. A detailed comparison of these two families of solutions  can be found  in\cite{beler-dereli1}. The odd-Grassmann  nature of the classical gravitino fields were later emphasized\cite{finkelstein-kim} and elaborated on\cite{volkov-galtsov}. A generalization with non-vanishing gravitino energy-momentum densities was made\cite{embacher1} and the relevance of Killing spinors for supersymmetric families of solutions was noted\cite{hull}. In all these papers,
the space-time metric is given by the Ehlers-Kundt pp-wave metric
\ba
g = 2 dU \otimes dV + 2 d\zeta \otimes d\bar{\zeta} + 2 H(U,\zeta,\bar{\zeta}) dU \otimes dU
\ea
in Brinkmann coordinates
$$
U= \frac{1}{\sqrt{2}}(z+t), \quad V =  \frac{1}{\sqrt{2}}(z-t) , \quad \zeta =  \frac{1}{\sqrt{2}}(x+iy), \quad \bar{\zeta} =  \frac{1}{\sqrt{2}}(x-iy).
$$
However,the usual practice to study the interactions of a gravitational pp-wave with test masses involves transforming to Rosen coordinates
\ba
U=u, \; V=v- \frac{x^2}{2}F(u)F^{\prime}(u) -\frac{y^2}{2}G(u)G^{\prime}(u) , \; X=F(u)x, \; Y=G(u)y ,
\ea
so that the pp-wave metric becomes
\ba
g = 2 du \otimes dv + F^{2}(u) dx \otimes dx + G^{2}(u) dy \otimes dy ,
\ea
provided $2H(u,x,y) = x^2 F F^{\prime \prime} + y^2 G G^{\prime \prime}$.  The transformation of the gravitino ansatz in Brinkmann  coordinates  to an expression in Rosen coordinates is non-trivial.
The explicit set of $\gamma$-matrices in the Majorana realization that are used to express the gravitino ansatz in Brinkmann coordinates needs to be switched
to a $\gamma$-matrix set that would still provide a Majorana realization but now relative to the new orthonormal co-basis in Rosen coordinates.
This is not a straightforward task.  The approach we take in this paper is to start with the pp-wave metric in Rosen coordinates.
Then we go back to the complex quaternion valued exterior differential forms over the 4-dimensional space-time \cite{dereli-tucker}.
Exterior differential forms are coordinate independent and all relevant tensors in this language remain free of indices $\mu,\nu,\dots$ in a coordinate chart $\{x^{\mu}\}.$
The algebra of complex quaternions on the other hand carry representations of the spin cover of the local Lorentz group, that is, $SL(2,{\mathcal{C}})$. As such it can be used
to express  the local frames, connection and curvature forms without use of Lorentz indices $a,b,\dots$.  Thus an algebraic formulation of 4-dimensional space-time geometries is achieved without any indices\cite{tucker}. The massless chiral spinor fields are introduced in this language by considering minimal left ideals as spin basis\footnote{These correspond to undotted spinors in  the NP formalism. In our approach dotted spinors of NP formalism are identified as minimal right ideals with a corresponding spin basis.}.The advantage of our unconventional formalism is that once a local Lorentz frame is chosen, it automatically fixes the corresponding spin basis, so that one doesn't have to worry about determining
a new set of Majorana matrices  that blends in well with the transformed co-frame. The derivation of the simple supergravity field equations in terms of complex quaternion valued exterior differential forms were given in detail before\cite{benn}.Similar techniques can also be found in \cite{morita2}.

\medskip

 \noindent The organization of the paper is as follows. In Section:2 we give a brief introduction to complex quaternion valued exterior differential forms over 4-dimensional space-time manifolds. Local Lorentz connection 1-forms are given  and the corresponding torsion and curvature 2-forms are determined from the Cartan structure equations.
 We discuss  the spin bases and their properties in the algebra of complex quaternions. We work in particular with left-chiral massless spinors.
 Next we give simple supergravity field equations in their final form in Section:3.  We will talk about an action density 4-form in order to comment on the local supersymmetry covariance of the field equations and in view of the conservation laws related with the symmetries of the action. However, we will refrain from any variational derivation and refer to previous work.
Section:4 is devoted to the discussion of pp-wave ansatz for the metric and the gravitino field in Rosen coordinates. We substitute these in to the coupled field equations and reduce them to a single second order ordinary differential equation satisfied by the metric function. In case a gravitino distribution that describes a progressive wave is specified,
 the metric function can be explicitly determined.Section:5 is reserved for concluding comments.

\bigskip


\section{Complex Quaternionic Differential Forms}
Any element of the algebra of complex quaternions $\mathbb{C}\otimes\mathbb{H}$ can be written as a real linear combination of the following eight elements $1$, $i$, $\hat{e_1}$, $\hat{e_2}$, $\hat{e_3}$, $i\hat{e_1}$, $i\hat{e_2}$ and $i\hat{e_3}$. The complex unit $i \in \mathbb{C}$ commutes with every element of the algebra with $i^2=-1$ and $\hat{e_1}$, $\hat{e_2}$, $\hat{e_3}$ are the quaternionic units denoted by $\hat{e_k} \in \mathbb{H}$ such that they obey the commutation relations

\begin{equation}
 \hat{e_k}\hat{e_j}=-\delta_{kj}+\epsilon_{kjl}\hat{e_l}.
\end{equation}
One can easily identify $i\hat{e_1}$, $i\hat{e_2}$ and $i\hat{e_3}$, respectively, with the usual Pauli matrices $\sigma_1$, $\sigma_2$ and $\sigma_3$ which are the three objects that generate rotations; while the multiplication and commutation rules of $\hat{e_1}$, $\hat{e_2}$, $\hat{e_3}$ similarly demonstrate that they are the source of boosts. Finally, these six items may be proven to produce the Lorentz algebra. We would get an element of $SL(2,\mathbb{C})$ if we took a real linear combination of these six items, multiplied by complex $i$ and exponentiated the whole thing. Complex conjugation is defined by the map $(i \rightarrow -i)$ and denoted by a superscript $^*$, while quaternionic conjugation $(\hat{e_k}\rightarrow - \hat{e_k})$ is indicated by an overbar. Their composition is denoted by $^\dagger$.
Elements  $Q \in SL(2,{\mathcal{C}})$ are unit norm quaternions such that $Q\bar{Q}=\bar{Q}Q=1$.  An explicit realization is given by
$$
Q=e^{\frac{i}{2}\hat{a} \alpha} e^{\frac{1}{2}\hat{b} \beta}
$$
where $\hat{a}$ and $\hat{b}$ are unit q-vectors and $\alpha,\beta$ are real parameters. In general given a local Lorentz transformation $Q = q_4 + \sum q_k \hat{e}_k \in SL(2,\mathcal{C})$, ona may as well write
$$
\tilde{Q} = \left (\begin{array}{cc} q_4 -iq_3 & -q_2 -iq_1 \\ q_2-iq_1 & q_4+iq_3 \end{array} \right )
$$
 so that $Q\bar{Q}=\bar{Q}Q=1$ if and only if $det{\tilde{Q}}=1$. (We show both $Q$ and $\tilde{Q}$ by the  same symbol in what follows without a cause for confusion.)

\smallskip

\noindent Take any complex differential $p$-form $A$. It may be written in terms of the elements $\hat{e_k}$ by
\begin{equation}
  A^{(p)}=A^{(p)}_4 + \sum_{k=1}^{3} A^{(p)}_k \hat{e_k},
\end{equation}
where $(A_4,A_1,A_2,A_3)$ are complex valued $p$-forms.

\smallskip

\noindent The following operators working on arbitrary quaternions $q$ are defined in terms of complex and quaternionic conjugation as
\ba
2Re(q)=q+q^*, \quad 2Im(q) = q - q^*, \nonumber \\  \quad 2Sc(q)=q+\bar{q}, \quad
2Vec(q)=q-\bar{q}, \nonumber \\ 2\mathcal{H}(q)=q+q^{\dagger}, \quad 2\mathcal{A}(q)=q-q^{\dagger}.
\ea
\noindent
More precisely, $q$ is termed a $q$-scalar for $Vec(q)=0$, a $q$-vector when $Sc(q)=0$, Hermitian and anti-Hermitian when $\mathcal{H}(q)=q$ and $\mathcal{H}(q)=-q$,respectively.

\medskip

\noindent  The Weyl spinors of $SL(2,\mathcal{C})$ on the other hand are represented by 2-component complex vectors transforming as
\ba
\phi_{\alpha} \rightarrow \tilde{\phi}_{\alpha} = (Q\phi)_{\alpha} , \quad  \phi^{\alpha} \rightarrow {\tilde{\phi}}^{\alpha}= (\phi \bar{Q})^{\alpha}
\ea
for undotted spinors and
\ba
\psi_{\dot{\alpha}} \rightarrow {\tilde{\psi}}_{\dot{\alpha}}= (\psi Q^{\dagger})_{\dot{\alpha}} , \quad  {\psi^{\dot{\alpha}}} \rightarrow {\tilde{\psi}}^{\dot{\alpha}} = (Q^{*}\psi)^{\dot{\alpha}}
\ea
for dotted spinors.
These Weyl spinors can be represented  in the algebra of complex quaternions  by left ideals generated by $L_1:(U^1,U^2)$, $L_2:(W^1,W^2)$
or the right ideals generated by $R_1:(U^1,W^2)$ and $R_2:(U^2,W^1)$
where
\ba
U^1=\frac{1}{\sqrt{2}}(1+i\hat{e_3}), \quad U^2=\frac{1}{\sqrt{2}}(\hat{e_2}+i\hat{e_1}), \nonumber \\
W^1=\frac{1}{\sqrt{2}}(1-i\hat{e_3}), \quad W^2=\frac{1}{\sqrt{2}}(\hat{e_2}-i\hat{e_1}).
\ea
The multiplication table for the elements of these spin bases is given below. (An entry in the first row is  multiplied from the left by an entry in the first column.)

\medskip

\begin{center}
\begin{tabular}{l|cccc}
& $U^1$ & $U^2$ & $W^1$ & $W^2$ \\
\hline
 $U^1$ &  $\sqrt{2}U^1$ & $0$   &  $0$  & $\sqrt{2}W^2$ \\
 $U^2$ &  $\sqrt{2}U^2$  &  $0$  &  $0$  & $-\sqrt{2}W^1$  \\
 $W^1$ &   $0$   & $\sqrt{2}U^2$   &  $\sqrt{2}W^1$  &  $0$ \\
 $W^2$ &    $0$  &$-\sqrt{2}U^1$    & $\sqrt{2}W^2$   & $0$ \\
\end{tabular}
\end{center}

\bigskip

\noindent  An $SL(2,\mathbb{C})$ transformation applied to spinors can be induced by the left and right action of a unit quaternion $Q$ ($Q\bar{Q}=\bar{Q}Q=1$) according to
\begin{alignat}{4}
\phi&=\phi_1U^1+\phi_2U^2 &&\rightarrow&& \hspace{2mm} Q\phi \nonumber \\
\dot{\chi}&=\chi_{\dot{1}}W^1+\chi_{\dot{2}}W^2 &&\rightarrow&& \hspace{2mm} Q^{*}\dot{\chi} \nonumber \\
\psi&=\psi_{1}U^1+\psi_{2}W^2 &&\rightarrow&& \hspace{2mm} \psi \bar{Q} \nonumber \\
\dot{\xi}&=\xi_{\dot{1}}U^2+\xi_{\dot{2}}W^1 &&\rightarrow&& \hspace{2mm} \dot{\xi} Q^{\dagger}.
\end{alignat}

\medskip

\noindent A Majorana spinor valued p-form is one where all four types of complex spinor p-forms can be expressed in terms of two independent complex spinor p-forms.
Suppose they are given by $\phi_A$ and $\phi_B$. Then
\begin{eqnarray}
\phi_{+} &=& -\phi_B U^1 + \phi_{A} U^2, \quad  \phi_{-} = \phi_A W^1 + \phi_{B} W^2, \nonumber \\ {\dot{\phi}}_{+} &=&  i \left ( \phi_B^{*} U^1 + \phi_{A}^{*} W^2 \right ),
\quad  {\dot{\phi}}_{-} = i \left ( \phi_{A}^{*}  W^1-\phi_{B}^{*} U^2 \right ).
\end{eqnarray}
In general, Majorana spinor valued p-forms satisfy
\ba
{\phi^{\dagger}_{\pm}}= \pm i {\dot{\phi}}_{\pm} .
\ea

\medskip

\noindent We give below a brief description of the 4-dimensional space-time geometry
in the language of complex quaternion valued exterior differential forms.  Let us first consider the (anti-Hermitian) co-frame 1-form
\begin{equation}
  e=ie^0+\sum_{k=1}^{3}e^k\hat{e_k}=-e^{\dagger}
\end{equation}
in terms of which we can express the metric of spacetime as
\ba
g=Re(e\otimes\bar{e}),
\ea
The usual tetrad components $e^{a}_{\mu}$ are defined by
\begin{equation}
  e^a=e^a_{\mu}dx^{\mu} \quad (a=0,1,2,3)
\end{equation}
in a local coordinate chart $\{x^{\mu}\}$.
An $SL(2,\mathbb{C})$ transformation converts the co-frame according to
\begin{equation}
  e \rightarrow \hspace{2mm} QeQ^{\dagger}.
\end{equation}

\noindent
The following are the definitions for the connection, torsion, and curvature forms over space-time:
\ba
\hat{\omega}&=&\sum_{k=1}^{3}\omega^k\hat{e_k}, \nonumber \\
T&=&de + \hat{\omega}\w e + e \w \hat{\omega}^{\dagger}=iT^0+\sum_{k=1}^{3}T^k\hat{e_k}=-T^{\dagger}, \nonumber \\
\hat{R}&=&d\hat{\omega}+\hat{\omega} \w \hat{\omega} = \sum_{k=1}^{3}R^k\hat{e_k},
\ea
with\footnote{There is no sign inconsistency here because the spatial indices can be raised or lowered without sign change.}
\ba
\omega^k=-\frac{1}{2}(i\omega^0_{\;\;k}+\frac{1}{2}\epsilon_{ijk}\omega^{i}_{\;\;j}), \quad
R^k=-\frac{1}{2}(iR^0_{\;\;k}+\frac{1}{2}\epsilon_{ijk}R^{i}_{\;\;j})
\ea
where $\omega_{ab}=-\omega_{ba}$ are the real components of the connection 1-forms and where $i,j,k, . . . = 1,2,3$ and the symbol $\epsilon_{ijk}$ is totally antisymmetric with $\epsilon_{123} = +1$. The frame indices are raised and lowered through the Minkowski metric $\eta_{ab}=diag(-,+,+,+)$. It should be noted from above that $T$ is anti-Hermitian so the components $(T^0,T^1,T^2,T^3)$ are real 2-forms. Moreover $\hat{\omega}$ and $\hat{R}$ are $SL(2,\mathbb{C})$ valued; consequently $\omega^k$ and $R^k$ are complex 1-forms.
We note the Bianchi identities that follow from the structure equations above as their integrability conditions:
\ba
dT + \hat{\omega} \wedge T -T \wedge \hat{\omega}^{\dagger} = \hat{R} \wedge e - e \wedge {\hat{R}}^{\dagger},
\ea
and
\ba
d\hat{R} + \hat{\omega} \wedge \hat{R} - \hat{R} \wedge \hat{\omega} = 0.
\ea
It is not difficult to verify the following local Lorentz transformation rules
$$
\hat{\omega} \rightarrow Q \hat{\omega} \bar{Q} + Qd\bar{Q}, \quad  T \rightarrow O T Q^{\dagger}, \quad \hat{R} \rightarrow Q \hat{R} \bar{Q}.
$$
Consider a Weyl spinor that transforms as $\phi \rightarrow Q\phi$. Then we define its  covariant exterior derivative
$$
\nabla \phi = d\phi + \hat{\omega} \phi
$$
which transforms properly under a local Lorentz transformation as $\nabla \phi \rightarrow  Q \nabla \phi$.  Similary for a dotted spinor we have
$$
\nabla\dot{\phi} = d \dot{\phi} - \dot{\phi} {\hat{\omega}}^{\dagger}.
$$

\medskip

\noindent We now outline the essentials of complex null basis 1-forms  because we will be using them extensively in our computations.
First and foremost, the relationships between the orthonormal and complex null co-frames are fixed as
\ba
l=\frac{1}{\sqrt{2}} (e^3+e^0), \quad n=\frac{1}{\sqrt{2}} (e^3-e^0), \quad m=\frac{1}{\sqrt{2}}(e^1+ie^2 ).
\ea
We can re-write  the metric as
\begin{equation}
  g=l \otimes n +n \otimes l  + m \otimes m^{*} + m^{*} \otimes m;
\end{equation}
and the orientation of spacetime is fixed by $*1=e^0 \w e^1 \w e^2 \w e^3 =i l \w n \w m \w m^{*}$, where $*$ to the left denotes the Hodge dual defined on forms.
One should take care that a ${*}$ to the right of a symbol on the other hand denotes complex conjugation.
We switch to a more convenient  notation at this point by setting
\ba
\omega_{\pm}=\omega^1\pm i\omega^2,\quad \omega_0=\omega^3, \quad
R_{\pm}=R^1\pm iR^2, \quad  R_0=R^3.
\ea
It is now possible to expand explicitly in terms of null tetrads and  the spinor basis elements in the following way:
\ba
e&=&iW^1l -iU^1n-iU^2m+iW^2m^{*}, \nonumber \\
\hat{\omega}&=&\frac{i}{\sqrt{2}} \omega_{-} \; W^2 -\frac{i}{\sqrt{2}} \omega_{0} \; (U^1-W^1) -\frac{i}{\sqrt{2}} \omega_{+} \; U^2 ,  \nonumber \\
\hat{R}&=&\frac{i}{\sqrt{2}} R_{-} \; W^2- \frac{i}{\sqrt{2}}R_{0}\;(U^1-W^1) -\frac{i}{\sqrt{2}}R_{+} \; U^2.
\ea

\section{Simple Supergravity Field Equations}
The basic field variables of the theory consists of a co-frame field $e$ in terms of which the Lorentzian metric
\ba
g= Re(e \otimes \bar{e})
\ea
and a gravitino 1-form $\chi$ that is an (odd-Grassmann valued)  Majorana spinor 1-form such that
\ba
\chi^{\dagger} = -i \dot{\chi}.
\ea
The connection 1-form $\hat{\omega}$ is also treated as an independent variable in our first order variational formalism.
We are going to vary the action $$I[e,\hat{\omega},\chi]=\int_{M} \mathcal{L}$$
where the Lagrangian density 4-form
\ba
{\mathcal{L}} = ImSc \left ( k \hat{R} \wedge e \wedge e^{*}  -i\bar{e} \wedge (\chi \wedge \nabla{\dot{\chi}}) -i\bar{e} \wedge (\nabla \chi \wedge \dot{\chi} )  \right ) .
\ea
$k=\frac{1}{8\pi G}$ is the universal gravitational coupling constant.
It is not difficult to verify  that under an infinitesimal local supersymmetry transformation given by
\ba
\delta e = 2 \mathcal{A}( i\epsilon \dot{\chi}), \quad \delta \chi = k\nabla \epsilon, \quad  \delta \dot{\chi} = k\nabla \dot{\epsilon},
\ea
where the supersymmetry parameter $\epsilon$ is a Majorana spinor, the Lagrangian density 4-form above changes by a closed form.
The first order variations of the above action yields the following
 simple supergravity field equations:
\ba
\hat{R} \wedge e = \frac{i}{k} \nabla \chi \wedge \dot{\chi},
\ea
\ba
\bar{e} \wedge \nabla \chi =0 ,\quad
\ea
\ba
T = \frac{i}{k} \chi \wedge \dot{\chi}.
\ea
It should be noted that there is more to the Einstein field equations than  given above by Eqn.(29). In fact
\ba
-2\mathcal{H}(\hat{R} \wedge e) \equiv i G = G_0+\sum_{k=1}^{3}G_k(i\hat{e}_k)
\ea
give us the Einstein 3-forms
\ba
G_a = -\frac{1}{2} R^{bc} \wedge {}^{*}e_{abc} .
\ea
 Similarly
 \ba
 -2\mathcal{H}(i \nabla \chi \wedge \dot{\chi}) \equiv i \tau = \tau_0+\sum_{k=1}^{3}\tau_k (i\hat{e}_k)
 \ea
 give us the gravitino stress-energ-momentum 3-forms
 \ba
 \tau_a = T_{ab} {}^{*}e^b.
  \ea
 In order to present the Einstein field equations as given by Eqn.(29), one should employ the Bianchi identities (19) and (20).


\section{Exact pp-Wave Solutions}
Consider  gravitational pp-wave spacetimes with the metric
\ba
g = 2 du \otimes dv + F^{2}(u) dx \otimes dx + G^{2}(u) dy \otimes dy
\ea
given in Rosen coordinates $\{u,v,x,y\}$. Then we introduce the complex null basis 1-forms
\ba
l = du, \quad n = dv , \quad  m = \frac{1}{\sqrt{2}}(F(u)dx + iG(u)dy) .
\ea
The corresponding Levi-Civita connection 1-forms are determined from the Cartan structure equations
\ba
de +\hat{\Gamma} \wedge e +e \wedge {\hat{\Gamma}}^{\dagger} = 0
\ea
as follows:
\ba
\Gamma_0 = 0, \quad \Gamma_{-} = -\frac{i}{2} (\frac{F^{\prime}}{F} - \frac{G^{\prime}}{G})m - \frac{i}{2
} (\frac{F^{\prime}}{F} +\frac{G^{\prime}}{G}) m^{*} , \quad \Gamma_{+} = 0 ,
\ea
where ${}^{\prime} \equiv \frac{d}{du}.$
The gravitino field is given by the Weyl spinor valued 1-form
\ba
\chi = \varphi_{1}(u) m U^{1}
\ea
where $\varphi_{1}(u)$ is an odd-Grassmann valued complex function of $u$ only. We have the Hermitian conjugate gravitino
\ba
\chi^{\dagger} = \varphi_{1}^{*}(u) m^{*} U^1 .
\ea
A few comments are in order here. It is possible to introduce a time coordinate in terms of  null coordinates
\ba
u = \frac{1}{\sqrt{2}} (z+t) , \quad v = \frac{1}{\sqrt{2}} (z-t).
\ea
Then the metric functions that depend only on $u$ describe progressive waves that move along the negative $z$-axis.They are left-movers.
It is possible to describe right-movers on the other hand by taking  metric functions that depend on $v$ only.
This other case can be treated similarly  where the gravitino ansatz now reads
\ba
\chi = \varphi_{2}(v) m^{*} U^2.
\ea
In what follows we deal only with left-movers.
We first work out the torsion 2-form
\ba
T = -\sqrt{2} \varphi_{1}(u) \varphi_{1}^{*}(u) m \wedge m^{*} \; U^1
\ea
so that the contortion 1-forms are determined as follows
\ba
K_0 = \frac{|\varphi_1|^2}{2\sqrt{2}} l , \quad K_{-} = -\frac{|\varphi_1|^2}{\sqrt{2}} m^{*}, \quad K_{+}=0.
\ea
When all these expressions are  put in the Rarita-Schwinger equation, it is  satisfied only for the choice $F=G$,valid up to an arbitrary multiplicative constant that may set equal to unity.\footnote{One must have $\frac{F^{\prime}}{F} - \frac{G^{\prime}}{G} =0$ .}
We have
\ba
\nabla \chi = \left ( \varphi_{1}^{\prime}(u) + \frac{F^{\prime}}{F} \varphi_{1}(u) -\frac{i}{2\sqrt{2}} |\varphi_{1}|^2 \varphi_{1}(u) \right ) l \wedge m \; U^1 .
\ea
It should be remarked that
$$
|\varphi_{1}|^2 \varphi_{1} = \varphi_{1}^{*} (\varphi_{1}\varphi_{1}) =0
$$
vanishes due to the odd-Grassmann nature of the gravitino field.
Then we move on to the Einstein field equations with
\ba
\hat{R} \wedge e = \left ( -i\frac{F^{\prime \prime}}{F} -\frac{1}{\sqrt{2}} \frac{d}{du} (|\varphi_{1}(u)|^2) -\sqrt{2}  |\varphi_{1}(u)|^2 \frac{F^{\prime}}{F}  \right )
l \wedge m \wedge m^{*} \; U^1 ,
\ea
and
\ba
-\frac{1}{k}\nabla \chi \wedge \chi^{\dagger} = - \frac{1}{k} \left (  \varphi_{1}^{\prime} \varphi_{1}^{*} +   |\varphi_{1}(u)|^2 \frac{F^{\prime}}{F}  \right ) l \wedge m \wedge m^{*} \; U^1 .
\ea
We now consider the polar decomposition
\ba
\varphi_{1}(u) = |\varphi_{1}(u)| e^{i\alpha(u)}
\ea
For an arbitrary phase function the gravitino field describes a progressive wave with a modulated amplitude. The particular choice
$\alpha (u) = \pm \kappa u$ would  correspond to a gravitino plane wave with wave number $\kappa$.
We further assume
\ba
|\varphi_{1}(u)| = \frac{|\eta |}{F(u)} > 0
\ea
where $\eta$ is a complex, odd-Grassmann constant.
Then the Einstein field equations
 are satisfied provided the following second order o.d.e is solved:
\ba
F^{\prime \prime} = |\eta|^2 \frac{ \alpha^{\prime}}{F}.
\ea
It is not difficult to verify that the expressions
\ba
F(u) = \sqrt{\frac{\kappa}{k}} \; |\eta | \; u \ln u , \quad \alpha (u) = -\kappa u +\kappa u \ln u .
\ea
give a non-trivial simple supergravity solution. Then the  gravitino energy-momentum 3-forms are found to be
\ba
i\tau = 2{\mathcal{H}}(-\nabla \chi \wedge {\chi}^{\dagger})  = -i \frac{2k}{u^2 \ln u} l \wedge m \wedge m^{*} U^1.
\ea
Therefore the non-vanishing components of the gravitino energy-momentum densities are
\ba
T_{00} = T_{33} = \frac{k}{u^2 \ln u} .
\ea



\bigskip

\section{Conclusion}

\noindent  A class of exact pp-wave solutions of simple supergravity field equations are constructed in Rosen coordinates. In fact the well-known gravitational pp-wave metric of Ehlers and Kundt is expressed in Brinkmann coordinates.The metric can be put in Rosen form by a coordinate transformation. The corresponding transformation of the gravitino 1-form
is rather involved and hasn't been worked out explicitly before. In this paper, we start instead directly with the pp-wave metric in Rosen coordinates and with a suitable ansatz for the gravitino 1-form, solve the simple supergravity field equations exactly. The powerful techniques of complex quaternionic exterior differential forms over 4-dimensional space-time are used  extensively in our construction. We gave a brief review of the formalism, mostly referring to our older work. The restriction  $F=G$  put on our metric functions by the
Rarita-Schwinger equation limits the advantages of going over to Rosen coordinates to discuss physical implications of the exact solutions we found.
Nevertheless, a family of non-gauge generated exact solutions of supergravity theories  may be of interest in its own right.

\bigskip

\section{Acknowledgement}
One of us (T.D.) thanks the Turkish Academy of Sciences (TUBA) for partial support.

\bigskip

\end{document}